\documentclass{article}
\usepackage{amsmath}
\usepackage{amssymb}
\usepackage{graphicx}
\usepackage{hyperref}
\usepackage{wrapfig}
\usepackage{bbding}

\newcommand{\D}{\,\mathrm{d}}
\newcommand{\pd}[2]{\ensuremath{\frac{\partial #1}{\partial #2}}}
\newcommand{\pdl}[2]{\ensuremath{\partial #1 / \partial #2}}

\newcommand{\cf}{\textit{cf.} }
\newcommand{\eg}{\textit{e.g.}, }

\newcommand{\rs}{\ensuremath{\rho^{(s)}}}

\newcommand{\vnb}{\ensuremath{\mathbf{v}_\text{n}}}
\newcommand{\vo}{\ensuremath{v_\text{1}}}
\newcommand{\vob}{\ensuremath{\mathbf{v}_\text{1}}}
\newcommand{\vt}{\ensuremath{v_\text{2}}}
\newcommand{\vtb}{\ensuremath{\mathbf{v}_\text{2}}}

\newcommand{\job}{\ensuremath{\mathbf{j}_\text{1}}}
\newcommand{\jtb}{\ensuremath{\mathbf{j}_\text{2}}}

\author{L.A.\,Melnikovsky$\,^{a,b}$
\\
$^a\,$Weizmann Institute of Science, Rehovot, Israel\\
$^b\,$P.L.\,Kapitza Institute for Physical Problems, Moscow, Russia
}
\title{Vortices in Andreev-Bashkin Superfluids}
\date{January 11, 2024}
\begin{document}
\maketitle
\begin{abstract}
Andreev-Bashkin entrainment makes the hydrodynamics of the binary superfluid solution particularly interesting. We investigate stability and motion of quantum vortices in such system.
\end{abstract}

\section{Introduction}
Seminal 1975 paper \cite{ab} by A.F.\,Andreev and E.P.\,Bashkin shattered many preconceptions and opened an entire new field of two-species superfluids. The BCS-type superfluidity of $^3$He subsystem in $^4$He-$^3$He solutions, which Andreev and Bashkin had in mind, is yet to be observed experimentally, but the effect they predicted is important for the description of a diverse range of systems, including pulsars \cite{sauls}, atomic gases \cite{becvortex,2vortex,sound}, binary magnon \cite{magnons} and coupled mass-magnon condensates \cite{he3}. In the present work, we analyze the properties of quantized vortices in a double superfluid. We show that it can bear three elementary vortex flavors. In contrast to the usual case, the single-quantum vortices may be, somewhat paradoxically, unstable (see the caption to Fig.\ref{lattice}). It would be interesting to compare our results with the instabilities and vortex dynamics recently investigated numerically~\cite{split,patrick}.

If we use the rest frame of the normal flow ($\vnb=0$), then the low-velocity expansion of the kinetic energy density for an isotropic Andreev-Bashkin superfluid is given by the expression \cite{ab}:
\begin{equation}
\label{Ekinetic}
E=\frac{\rs_{11}\vo^2 + 2\rs_{12}(\vob\cdot \vtb)+ \rs_{22}\vt^2}{2},
\end{equation}
where only the terms depending on the superfluid velocities \vob\ and \vtb\ are retained.\footnote{The vortex energy (see below) is dominated by the distant flow, which is essentially incompressible in liquids. This approximation may be violated in gaseous systems, where the densities are equally important variables.} These velocities are proportional to the gradients of the corresponding condensate phases $\phi_1$ and $\phi_2$:
\begin{equation}
\label{grad}
\begin{aligned}
\vob &= \frac{\hbar}{m_1} \nabla \phi_1,\\
\vtb &= \frac{\hbar}{m_2} \nabla \phi_2.
\end{aligned}
\end{equation}
As emphasized in \cite{ab}, the bare particle masses $m_1$ and $m_2$ enter these relations (\cf elementary electron charge entering Ginzburg-Landau equations).

The stability of a uniform state implies that the matrix of the superfluid densities is positive definite
\begin{equation}
\label{rho-matrix}
\boldsymbol{\rho}^{(s)}=
\begin{pmatrix}
\rs_{11} & \rs_{12} \\[.5em]
\rs_{21} & \rs_{22}
\end{pmatrix}
\succ 0.
\end{equation}
The off-diagonal elements $\rs_{12}=\rs_{21}$ of this symmetric matrix are responsible for the entrainment effect; the mass fluxes of individual solution species $\job$, $\jtb$ depend on both velocities:
\begin{equation}
\label{flux}
\begin{aligned}
\job & = \rs_{11}\vob + \rs_{12}\vtb,\\
\jtb & = \rs_{21}\vob + \rs_{22}\vtb.
\end{aligned}
\end{equation}

\section{Quantum vortex lines}
\label{sec-vortex}
Equations \eqref{grad}, which describe the potential flow, can be violated along Onsager-Feynman vortex line singularities \cite{khalat}. In a pure Landau superfluid, the single-valuedness of its only condensate wave function constrains the superfluid velocity circulation to a multiple of $2\pi\hbar / m$. In the Andreev-Bashkin superfluid, both velocities $\mathbf{v}_{1,2}$ may acquire non-zero quantized circulations along a closed path around a single vortex:
\begin{equation*}
\oint \mathbf{v}_{1,2} \cdot \D\mathbf{l} = 2\pi\hbar \frac{n_{1,2}}{m_{1,2}}.
\end{equation*}
The integer numbers $n_1$ and $n_2$ uniquely identify this vortex.

Consider a solitary, straight vortex line. It generates an axially symmetric velocity field:
\begin{equation}
\label{vortex-field}
v_{1,2}=\frac{\hbar}{r} \frac{n_{1,2}}{m_{1,2}},
\end{equation}
where $r$ is the distance from the vortex. By integrating Eq.\eqref{Ekinetic} we get the vortex energy per unit length:
\begin{equation}
\label{energy}
\everymath{\displaystyle}
\epsilon =
\int_\xi^L E \,2\pi r \D r =
\pi \hbar^2 \ln\left(L/\xi\right)
\,
\mathbf{n}^T
\boldsymbol{H}\,
\mathbf{n},
\quad
\boldsymbol{H}=
\begin{pmatrix}
\frac{\rs_{11}}{m_1^2} & \frac{\rs_{12}}{m_1 m_2} \\[1em]
\frac{\rs_{21}}{m_1 m_2} & \frac{\rs_{22}}{m_2^2}
\end{pmatrix},
\end{equation}
where the integral spans from the vortex core radius $\xi$ to some spatial extent $L$, such as the vortex length or inter-vortex spacing.\footnote{Particularly, if $L$ is interpreted as a proper penetration depth, then the present analysis remains mostly valid for electrically charged superfluids \cite{yan,sauls}.} Here $\mathbf{n}^T = (n_1,n_2)$ is a fictitious two-dimensional vector representing the vortex ``charge''.

\section{Stability}
\label{sec-stability}
The energy \eqref{energy} depends quadratically on the circulation; this is why the multi-quantum vortices in $^4$He are unstable with respect to decay into elementary single-quantum ones. This problem becomes more interesting in an Andreev-Bashkin superfluid.
\begin{figure}[t]
\vspace{-4mm}
\includegraphics[width=1\linewidth]{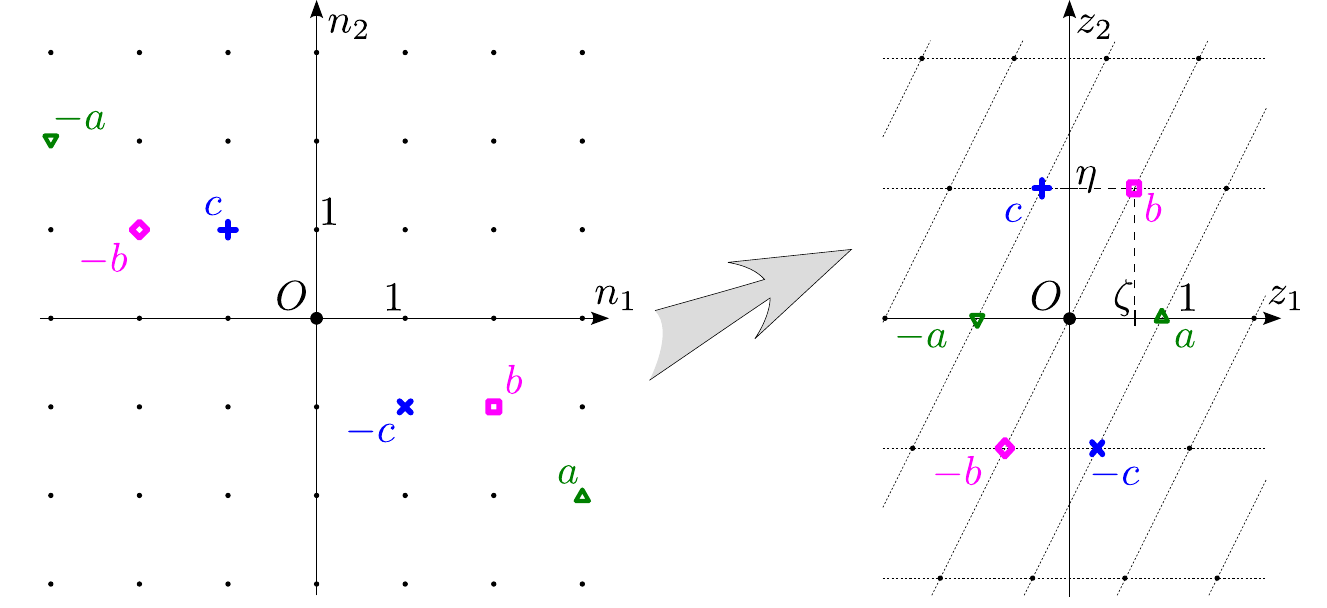}
\caption{Linear transformation of the fictitious vortex charge plane: the reduction of the energy quadratic form to a sum of squares distorts the square lattice. The figure corresponds to the energy quadratic form
$\epsilon\propto 38\,n_1^2+112\,n_1 n_2 + 85\,n_2^2$. In this example, the three stable vortex flavors are $\pm\mathbf{n}^{(a)}=\pm(3,-2)^T$, $\pm\mathbf{n}^{(b)}=\pm(2,-1)^T$, and $\pm\mathbf{n}^{(c)}=\pm(-1,1)^T$. One can readily see, \eg the energetic advantage of the decays $(1,0)^T \!\! \rightarrow b+c$ and  $(0,1)^T \!\!\rightarrow b+c+c$.}
\label{lattice}
\end{figure}

Spontaneous breakup\footnote{We only discuss the global stability here. The existence of metastable vortex configurations requires further investigation.} of a vortex $\mathbf{n}$ into two vortices $\mathbf{n}^{(1)}$ and $\mathbf{n}^{(2)}=\mathbf{n} - \mathbf{n}^{(1)}$ is possible if $\epsilon\left(\mathbf{n}\right) > \epsilon\left(\mathbf{n}^{(1)}\right) + \epsilon\left(\mathbf{n}^{(2)}\right)$. From Eq.\eqref{rho-matrix}, it follows that the quadratic form given by the matrix $\boldsymbol{H}$ in \eqref{energy} is also positive definite. It can be reduced to a sum of squares:
\begin{equation*}
\epsilon \propto
z_1^2 + z_2^2
\end{equation*}
by a linear transformation. This transformation deforms the integer square lattice of the vortex charges. Without loss of generality, we may assume that the shortest vector\footnote{The vector between two lattice nodes with the smallest Euclidean norm.} of the deformed lattice runs along the $Oz_1$ axis and has a unit length as shown in Fig.\ref{lattice}. All nodes of the new lattice can be parametrized by the integer vectors $\mathbf{k} \in \mathbb{Z}^2$:
\begin{equation*}
\mathbf{z}=
\begin{pmatrix}z_1\\z_2\end{pmatrix}=
k_1
	\begin{pmatrix}
	1\\0
	\end{pmatrix} + 
k_2
	\begin{pmatrix}
	\zeta \\ \eta 
	\end{pmatrix},
\end{equation*}
where $0 \le \zeta < 1$. A configuration $Z$ is unstable if and only if there is a node $Y$ such that $\angle ZYO > \pi/2$. It is obvious that only six\footnote{Analysis in higher dimensions is more cumbersome. In the $s$-dimensional case (for an $s$-species superfluid), there exist $2^s-1$ stable vortex flavors.} configurations (vortex and anti-vortex for each of three flavors $a$, $b$, and $c$) below are stable:
\begin{align*}
\mathbf{k}^{(a)}&=\pm
	\begin{pmatrix}
	1 \\ 0 
	\end{pmatrix},&\quad
\mathbf{k}^{(b)}&=\pm
	\begin{pmatrix}
	0 \\ 1 
	\end{pmatrix},&\quad
\mathbf{k}^{(c)}&=\pm
	\begin{pmatrix}
	-1 \\ 1 
	\end{pmatrix},\\
\text{or equivalently}\\
\mathbf{z}^{(a)}&=\pm
	\begin{pmatrix}
	1 \\ 0 
	\end{pmatrix},&\quad
\mathbf{z}^{(b)}&=\pm
	\begin{pmatrix}
	\zeta \\ \eta
	\end{pmatrix},&\quad
\mathbf{z}^{(c)}&=\pm
	\begin{pmatrix}
	\zeta-1 \\ \eta 
	\end{pmatrix}.
\end{align*}
Unfortunately, no universal closed-form expression for the stable vortex charges $\mathbf{n}^{(a)}$, $\mathbf{n}^{(b)}$, and $\mathbf{n}^{(c)}$ exists, but their identification for particular system parameters is straightforward using the Lagrange-Gauss reduction algorithm. High sensitivity of the stable charge vectors to the $\boldsymbol{\rho}^{(s)}$ matrix elements may result in complicated vortex branching structures for spatially inhomogeneous systems.

\section{Interaction energy}
\label{sec-interaction}
For multiple vortices of arbitrary shape, one can find the total energy by analogy~\cite{LL9} with the Biot-Savart formula for the magnetic field energy of linear currents:
\begin{equation}
\mathcal{E}=
\pi \hbar^2
\iint
	\mathbf{n}^{(1)T}\! \boldsymbol{H}\, \mathbf{n}^{(2)}
	\frac{\left(\D\mathbf{l}^{(1)} \cdot \D\mathbf{l}^{(2)}\right)}{2 r},
\end{equation}
where $r$ is the distance between two infinitesimal vortex fragments $\D\mathbf{l}^{(1)}$ and $\D\mathbf{l}^{(2)}$ with the charges $\mathbf{n}^{(1)}$ and $\mathbf{n}^{(2)}$. After integration, this gives the interaction energy (per unit length) of two straight, parallel vortices separated by a distance~$r$:
\begin{equation}
\epsilon=
2 \pi \hbar^2
\ln\left(L/r\right)
\mathbf{n}^{(1)T}\! \boldsymbol{H}\, \mathbf{n}^{(2)}
\quad\propto\quad
\mathbf{z}^{(1)}\!\cdot \mathbf{z}^{(2)}\, \ln\left(L/r\right)
.
\end{equation}
Qualitatively, the nodes forming an acute angle at the origin $O$ on the deformed $O z_1 z_2$ plane correspond to the vortices that repel each other (if some dissipation mechanism which reduces the energy is taken into account). And vice versa, the nodes at an obtuse angle correspond to the attracting vortices.

\section{Rotation}
\label{sec-rotation}
\begin{wrapfigure}[26]{r}{0pt}
\includegraphics[width=.36\linewidth]{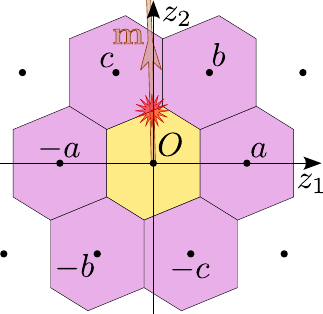}
\caption{Assume the particle masses in the system described in Fig.\ref{lattice} differ by a factor of two $m_2=2 m_1$. The respective mass vector $\mathbf{m}$ points approximately ``upward'' on the deformed vortex charge plane:
\mbox{$\mathbf{m}\propto (1,2)^T = 3\mathbf{n}^{(b)}+5\mathbf{n}^{(c)}$}.
At the critical point \SixteenStarLight, the $\mathbf{m}$-ray meets the boundary of the first Wigner-Seitz cell for the deformed lattice. In this particular example, the critical vortex is $\mathbf{n}_\text{cr}=\mathbf{n}^{(c)}$.}
\label{ws}
\end{wrapfigure}

Equilibrium in a cylindrical vessel rotating with angular velocity $\Omega$ delivers the minimum of the ``reduced'' energy $\mathcal{E}-M\Omega$, where $M$ is the angular momentum. The angular momentum of a single straight vortex (per unit length) positioned at the axis of a cylindrical vessel is easily obtained from \eqref{flux} and \eqref{vortex-field}:
\begin{equation}
\label{momentum}
\int_0^R 2\pi r \left(j_1 + j_2\right)
r \D r
=  \pi \hbar R^2
\mathbf{m}^T\! \boldsymbol{H}\, \mathbf{n}
,
\end{equation}
where $R$ is the vessel radius, and $\mathbf{m}^T = (m_1,m_2)$. The competition between the two expressions \eqref{momentum} and \eqref{energy} determines the threshold angular velocity $\Omega_\text{cr}$ and the flavor $\mathbf{n}_\text{cr}$ of the critical vortex which is spontaneously created upon reaching $\Omega_\text{cr}$. Geometrically, Eq.\eqref{momentum} means that the minimum of the reduced energy  ``travels'' along the mass vector $\mathbf{m}$, see Fig.\ref{ws}. The threshold corresponds to the intersection of this ray and the boundary of the first Wigner-Seitz cell for the deformed lattice.

At much higher rotation rates $\Omega \gg \Omega_\text{cr}$, the vortex distribution simulates (on average) the velocity field of rigid body rotation. The area concentrations of the circulation quanta for the two condensates are:
\begin{equation}
\label{feynman}
\nu_{1,2}=\frac{\Omega}{\pi\hbar}m_{1,2}
.
\end{equation}
Note that the relative concentration $\nu_1/\nu_2=m_1/m_2$ is generally irrational and can not be realized by any commensurate structure of $a$, $b$, and $c$ vortices. In other words, a special mass symmetry (\eg an approximate equality of the nucleon masses) is required for the periodic vortex lattice emergence in the Andreev-Bashkin superfluid. A potentially more significant fact for the problem of relative friction is that the area concentrations of real vortices $\nu_{a,b,c}$ may be higher than what Feynman's rule \eqref{feynman} predicts. For example, a particular choice of the numerical constants used in Figs.\ref{lattice},\ref{ws}, leads to $2\nu_1=\nu_2$ and $\nu_a=0$, $\nu_b=2\nu_1$, $\nu_c=3\nu_1$. 

\section{Kinematics}
\label{sec-motion}
According to Helmholtz's theorem~\cite{sonin}, the vortices in simple fluids are advected by the flow. Generalization to the Andreev-Bashkin superfluid is based on momentum conservation.

To avoid the difficulties and controversies related to the normal component~\cite{sonin}, assume the low-temperature limit and ignore the normal density. The momentum flux tensor~\cite{ab} is then given by:
\begin{equation}
\label{stress-tensor}
\Pi^{lk}=
   \rs_{\alpha \beta} v_\alpha^l v_\beta^k
    + P \delta^{lk},
\end{equation}
where $P$ is the pressure, $\alpha$ and $\beta$ enumerate the species, and summation over repeated subscripts is assumed.

Instead of investigating the effect of an external flow on the vortex motion it is easier to use the frame of reference where the vortex is at rest and analyze the conditions imposed on the flow. Consider the stationary velocity distribution in a plane, perpendicular to a straight vortex line. To eliminate the pressure-related term from \eqref{stress-tensor} in the continuity equation $0=\pdl{\Pi^{ik}}{x^k}$, take the curl of both sides:
\begin{equation}
\label{continuity}
0=e^{lj}\pd{}{x^j}\pd{\Pi^{lk}}{x^k}
=
\rs_{\alpha \beta} v_\beta^k \, \pd{}{x^k}\, e^{lj}
 \pd{v_\alpha^l}{ x^j}
,
\end{equation}
where $e^{lj}$ is the two-dimensional Levi-Civita tensor. To simplify Eq.\eqref{continuity}, we use the properties  of the $\rs_{\alpha \beta} = \rs_{\beta \alpha}$ symmetry and of the flow incompressibility $\pdl{v_\beta^k}{x^k}=0$, as well as the identity which follows from them:
\begin{equation*}
\rs_{\alpha \beta} \, e^{lj}\pd{v_\beta^k}{x^j} \pd{v_\alpha^l}{x^k}
=
\rs_{\alpha \beta} \, e^{lj}\pd{v_\beta^k}{x^j} 
	\left(\pd{v_\alpha^k}{x^l} + e^{lk}e^{pq}\pd{v_\alpha^p}{x^q}\right)=0.
\end{equation*}

The velocity is a superposition of the circular field generated by the vortex and an external wind produced by some other sources. We explicitly decompose the velocity as a sum $\mathbf{v}_\alpha + \mathbf{w}_\alpha$, where the first term is given by Eq.\eqref{vortex-field} and is due to the vortex (hence denoted $\mathbf{v}_\alpha$), and the second term $\mathbf{w}_\alpha$ is irrotational and designates the wind flow only. The equation \eqref{continuity} is obviously satisfied by the fields $\mathbf{v}_\alpha$ and $\mathbf{w}_\alpha$ separately, and the only remaining non-trivial term is:
\begin{equation}
\label{zero-magnus}
0=
\rs_{\alpha \beta} w_\beta^k \pd{}{x^k}\, e^{lj}
 \pd{v_\alpha^l}{ x^j}
 \quad\propto\quad
\sum\limits_{\alpha,\beta}
\frac{\rs_{\alpha \beta}}{m_\alpha} w_\beta^k n_\alpha 
=
\sum\limits_{\alpha}
j_\alpha^k \frac{n_\alpha}{m_\alpha}.
\end{equation}
Here the wind velocity $\mathbf{w}_\beta$ is taken at the vortex core (this is the only point where the curl is nonzero $e^{lj} \pdl{v_\alpha^l}{ x^j} \ne 0$).
Qualitatively, this corresponds to the cancellation of the Magnus forces\footnote{The total external force required to keep the vortex (per unit length) at rest in a moving fluid is given by:
\begin{equation*}
f^l=2\pi\hbar e^{lk} \sum\limits_{\alpha}
j_\alpha^k \frac{n_\alpha}{m_\alpha}.
\end{equation*}
This force may be provided, \eg by some pinning potential or an electric field acting on trapped charges.}
from the individual species. One can extract the vortex line velocity $\mathbf{v}_\text{L}$ in an arbitrary frame of reference from Eq.\eqref{zero-magnus}:
\begin{equation}
\label{vl}
\mathbf{v}_\text{L}=
\left(\frac{n_1}{m_1}\job + \frac{n_2}{m_2}\jtb\right)
\left/
\left(\frac{n_1}{m_1}\rho_1 + \frac{n_2}{m_2}\rho_2\right)\right.
,
\end{equation}
where $\rho_1=\rs_{11}+\rs_{12}$ and $\rho_2=\rs_{21}+\rs_{22}$ are the mass densities of the species, and the denominator can be conveniently expressed as $\mathbf{m}^T\! \boldsymbol{H}\, \mathbf{n}$, \cf Eq.\eqref{momentum}. A notable observation is that due to this denominator, the vortex velocity $v_L$ may exceed both $w_1$ and $w_2$. For example, assuming the numbers used in Figs.\ref{lattice},\ref{ws}, the $a$-vortex velocity is $\mathbf{v}_\text{L}^{(a)}=2\mathbf{w}_2-\mathbf{w}_1$.

Equation \eqref{vl} can be used to find the velocity of one vortex $\mathbf{n}^{(1)}$ in the field of the other $\mathbf{n}^{(2)}$:
\begin{equation}
v_\text{L}^{(1|2)}=
\frac{\hbar}{r}
\frac{\mathbf{n}^{(2)T}\! \boldsymbol{H}\, \mathbf{n}^{(1)}}{\mathbf{m}^T\! \boldsymbol{H}\, \mathbf{n}^{(1)}}
.
\end{equation}
The velocity of a circular vortex ring of radius $R_0$ is given by:
\begin{equation}
v_\text{ring}=\frac{\varkappa}{2R_0}\ln\frac{R_0}{\xi}=
\frac{\hbar}{2R_0}
\frac{\mathbf{n}^{T}\! \boldsymbol{H}\, \mathbf{n}}{\mathbf{m}^T\! \boldsymbol{H}\, \mathbf{n}}
\ln\frac{R_0}{\xi}
\end{equation}
and is easily obtained from the usual $^4$He expression \cite{LL9} by setting:
\begin{equation}
\varkappa = \hbar \frac{\mathbf{n}^{T}\! \boldsymbol{H}\, \mathbf{n}}{\mathbf{m}^T\! \boldsymbol{H}\, \mathbf{n}}.
\end{equation}
Similarly, the Kelvin wave dispersion is given by:
\begin{equation}
\omega=
\frac{\hbar k^2}{2}
\frac{\mathbf{n}^{T}\! \boldsymbol{H}\, \mathbf{n}}{\mathbf{m}^T\! \boldsymbol{H}\, \mathbf{n}}
\ln\frac{1}{k\xi}
.
\end{equation}

\section{Summary}
Quantized vortex in a double superfluid is a very interesting object. It is characterized by two integer quantum circulation numbers, which are conveniently combined into a single vector charge (Sec.\ref{sec-vortex}). The individual charge components couple due to the Andreev-Bashkin off-diagonal term. Only three distinct charge vectors correspond to energetically stable configurations (Sec.\ref{sec-stability}). These configurations depend on the superfluid densities matrix. All other vortices may spontaneously decompose into multiple elementary ones. The energy of a system of vortices can be expressed in terms of the pairwise interaction between individual vortex fragments (Sec.\ref{sec-interaction}). This interaction also results in the motion of one vortex fragment in the velocity field generated by the others (Sec.\ref{sec-motion}). An equilibrium rotating state of the fluid corresponds to a vortex distribution that is uniform on average but not generally periodic (Sec.\ref{sec-rotation}).

\section*{Acknowledgements}
I thank Sergey Kafanov, Vladimir Marchenko, Sam Patrick, Sivan Refaeli-Abramson, and Edouard Sonin for useful discussions.
This work was partially supported by the MOIA grant \#714481.
Present research would have been impossible without the protection by the IDF.
\eject


\begin{thebibliography}{99}
\bibitem{ab}
A.F.Andreev, E.P.Bashkin,
\textit{Three-velocity hydrodynamics of superfluid solutions},
Zh. Eksp. Teor. Fiz. \textbf{69}, 319 (1975)
[\href{http://jetp.ras.ru/cgi-bin/dn/e_042_01_0164.pdf}{JETP, \textbf{42}, 164 (1975)}].
\bibitem{sauls}
J.A.Sauls,
\textit{Superfluidity in the Interiors of Neutron Stars} in Timing Neutron Stars,
457 (Kluwer Academic Publishers, 1989)
[\href{https://arxiv.org/abs/1906.09641}{arXiv:1906.09641}].
\bibitem{becvortex}
M.R.Matthews, B.P.Anderson, P.C.Haljan, D.S.Hall, C.E.Wieman, and E.A.Cornell,
\textit{Vortices in a Bose-Einstein Condensate},
\href{https://doi.org/10.1103/PhysRevLett.83.2498}{Phys. Rev. Lett. \textbf{83}, 2498 (1999)}.
\bibitem{2vortex}
X.-C.Yao, H.-Z.Chen, Y.-P.Wu, X.-P.Liu, X.-Q.Wang, X.Jiang, Y.Deng, Y.-A.Chen, J.W.Pan,
\textit{Observation of Coupled Vortex Lattices in a Mass-Imbalance Bose and Fermi Superfluid Mixture},
\href{https://doi.org/10.1103/PhysRevLett.117.145301}{Phys. Rev. Lett. \textbf{117}, 145301 (2016)}.
\bibitem{sound}
J.H.Kim, D.Hong, Y.Shin,
\textit{Observation of two sound modes in a binary superfluid gas},
\href{https://doi.org/10.1103/PhysRevA.101.061601}{Phys. Rev. \textbf{A101}, 061601 (2020)}.
\bibitem{magnons}
P.Nowik-Boltyk, O.Dzyapko, V.E.Demidov, N.G.Berloff, S.O.Demokritov,
\textit{Spatially non-uniform ground state and quantized vortices in a two-component Bose-Einstein condensate of magnons},
\href{https://doi.org/10.1038/srep00482}{Sci Rep \textbf{2}, 482 (2012)}.
\bibitem{he3}
G.E.Volovik,
\textit{Spin Vortex Lattice in the Landau Vortex-Free State of Rotating Superfluids}, \href{https://doi.org/10.1134/S0021364020100045}{JETP Lett. \textbf{111}, 582 (2020)}.
\bibitem{split}
P.Kuopanportti, S.Bandyopadhyay, A.Roy,  D.Angom,
\textit{Splitting of singly and doubly quantized composite vortices in two-component Bose-Einstein condensates},
\href{https://doi.org/10.1103/PhysRevA.100.033615}{Phys. Rev. \textbf{A100}, 033615 (2019)}.
\bibitem{patrick}
S.Patrick, A.Gupta, R.Gregory, C.F.Barenghi,
\textit{Stability of quantized vortices in two-component condensates},
\href{https://doi.org/10.1103/PhysRevResearch.5.033201}{Phys. Rev. Res. \textbf{5}, 033201 (2023)}.
\bibitem{khalat}
I.M.Khalatnikov,
\textit{An Introduction to the Theory of Superfluidity}
(W.A.Benjamin, New York-Amsterdam, 1965).
\bibitem{yan}
G.A.Vardanyan, D.M.Sedrakyan,
\textit{Magnetohydrodynamics of superfluid solutions},
Zh. Eksp. Teor. Fiz. \textbf{81}, 1731 (1981)
[\href{http://www.jetp.ras.ru/cgi-bin/dn/e_054_05_0919.pdf}{JETP \textbf{54}, 919 (1981)}].
\bibitem{LL9}
E.M.Lifshitz, L.P.Pitaevsky,
\textit{Statistical Physics, Part 2}
(Pergamon Press, 1980), Sec 29.
\bibitem{sonin}
E.B.Sonin,
\href{https://doi.org/10.1017/CBO9781139047616}{\textit{Dynamics of Quantised Vortices in Superfluids}},
(Cambridge University Press, 2016).
\end{thebibliography}
\end{document}